\documentclass{ringb99}      
\usepackage{graphics}
\def\ros{{\sl ROSAT }}
\def\asca{{\sl ASCA }}

\def\arcmin{\ifmmode ^{\prime}\else$^{\prime}$\fi}
\def\arcsec{\ifmmode ^{\prime\prime}\else$^{\prime\prime}$\fi}
\def\approxlt{\mathrel{\hbox{\rlap{\lower.55ex \hbox {$\sim$}}
        \kern-.3em \raise.4ex \hbox{$<$}}}}
\def\approxgt{\mathrel{\hbox{\rlap{\lower.55ex \hbox {$\sim$}}
        \kern-.3em \raise.4ex \hbox{$>$}}}}

\begin{document}
\title{The central radio source in the NGC\,383 group: \\ 
       jet - IGM interaction ? }
\author{Stefanie Komossa, Hans B\"ohringer } 
\institute{Max--Planck--Institut f\"ur extraterrestrische Physik,
 Giessenbachstra{\ss}e, 85740 Garching, Germany \\
 {\sl email: skomossa@xray.mpe.mpg.de}  }
\authorrunning{St. Komossa, H. B\"ohringer}
\titlerunning{~ Jet-IGM interaction in the NGC\,383 group ?}
\maketitle

\vspace*{-8cm}
\begin{verbatim}
To appear in the proceedings of the Ringberg workshop on
`Diffuse Thermal and Relativistic Plasma in Galaxy Clusters'
(Ringberg castle, April 19-23, 1999); MPE Report.
\end{verbatim}

\vspace*{5.7cm}

\begin{abstract}

The NGC\,383 group is a rich group of galaxies that harbours the central
radio source 3C\,31 = NGC\,383. 
After presenting results from the X-ray analysis of the
extended intra-group medium and the central radio galaxy,
we discuss the spatial morphology of X-ray emitting gas 
as compared to the radio plasma  
and perform a comparison of the pressures 
of both components.
\end{abstract}

\section{Introduction}

The rich group of galaxies around NGC\,383 at redshift $z$=0.017 is 
located in the Perseus Pisces
filament (e.g., Zwicky et al. 1961, Arp 1966, Sakai et al. 1994).
The brightest member galaxy, NGC\,383 = 3C\,31, is a moderately bright radio galaxy
and has been extensively studied at radio wavelengths in the past (e.g.,
Klein \& Wielebinski 1979, Ekers et al. 1981, Condon et al. 1991, Artyukh et al. 1994,
Lara et al. 1997).
It shows a symmetric edge darkened
double-source structure with two strong jets.
The origin of the 
structures seen in the radio morphology of
dominant cluster/group members is still not well understood.
The structures of the jets of NGC\,383 were interpreted
by Blandford \& Icke (1978) as arising from the tidal interaction with the companion
galaxy NGC\,382. However, based on optical data
Fraix-Burnet et al. (1991) found no evidence
for interaction between the two galaxies.

Here, we present an analysis of the X-ray properties of NGC\,383,
the extended intra-group medium (IGM), and a search for radio-X-ray
relations on the basis of our {\sl ROSAT} PSPC observations.
These data presented here have been previously partly analyzed by
Trussoni et al. (1997; T97 hereafter) in a study of hot coronae in nearby
radio galaxies. A re-analysis of the \ros data and new \asca results 
have been recently presented by Hwang et al. (1999). For
details on the present study, see Komossa \& B\"ohringer (1999).

Physical parameters are calculated for $H_{\rm o}$ = 50 km/s/Mpc
and $q_{\rm o}$ = 0.5.
For the distance of the NGC\,383 group, 1\arcsec~corresponds to a scale of 0.5 kpc.

\section{X-ray analysis of the intra-group medium}

Widely extended X-ray emission
is present (Fig. 1).
Its spectrum is best described by thermal `Raymond \& Smith' emission
with a temperature of $kT \simeq$ 1.5 keV (using metal abundances
of 0.3 $\times$ solar), 
approximately constant in the measured region.
The total (0.1-2.4 keV) X-ray luminosity is $L_{\rm x} = 1.5\,10^{43}$ erg/s.

The extended X-ray emission can be traced out to a distance of
$\sim$ 33\arcmin, corresponding to $\sim$ 1 Mpc.
Fitting a standard `$\beta$-model'
of the form
\begin{equation}
 S = S_0 (1+ {r^2 \over {r_{\rm c}^2}})^{-3\beta + {1 \over 2}}
\end{equation}
to the azimuthally averaged surface brightness profile yields a slope
parameter $\beta = 0.38$ and a core radius $r_{\rm c} = 73$ kpc.
The gas mass enclosed inside 1 Mpc amounts to $M_{\rm gas}$ = 1.3\,10$^{13}$ M$_{\odot
}$.
Assuming spherical symmetry and the group to be approximately in hydrostatic
equilibrium,
the total gravitating mass 
is given by
\begin{equation}
M_{\rm total}(r) = -{k \over {\mu m_{\rm p} G}}\,T(r)\,r\,({r \over T}{dT \over dr} +
{r \over \rho}{d\rho \over dr})~~.
\end{equation}
We derive an integrated total mass of $M_{\rm total}$ = 6.3\,10$^{13}$ M$_{\odot}$
 within 1 Mpc
and a gas mass fraction of 21\%.
The profile of total and gas mass is displayed in Fig. 2.

The cooling time of the gas in the group center
is $t \simeq 2.7\,10^{10}$ yr, i.e. no `large-scale'
cooling flow is expected to have developed.
Further, the enhanced X-ray emission from the direction
of NGC 383 is consistent with originating from a point source.

\begin{table}[h]
 \caption{Summary of the properties of the NGC\,383 group of galaxies 
  derived from
  the X-ray analysis.}
\normalsize
     \label{gprop}
      \begin{tabular}{l}
      \hline
      \noalign{\smallskip}
spectral fit: \\
\indent ~~ $kT$=1.5 keV, $L_{\rm x}^{\rm 0.1-2.4 keV}$ = 1.5\,10$^{43}$ erg/s\\
beta-model results:  \\
\indent ~~ $S_{\rm 0} = 2.7\,10^{-3}$ cts/s/arcmin$^2$, $\beta = 0.38$,
$r_{\rm c} = 73$ kpc \\
central density, mass: \\
\indent ~~ $n_0$ = 1.3\,10$^{-3}$ cm$^{-3}$; $M_{\rm total}$ = 0.6\,10$^{14}$ M$_{\odot}$, \\
\indent ~~ gas mass fraction 21\% (at $r$ = 1\,Mpc) \\
      \noalign{\smallskip}
      \hline
 \end{tabular}
   \end{table}

\begin{figure*}[ht]
\vspace{-1.8cm}
 \resizebox{\hsize}{!}{\includegraphics{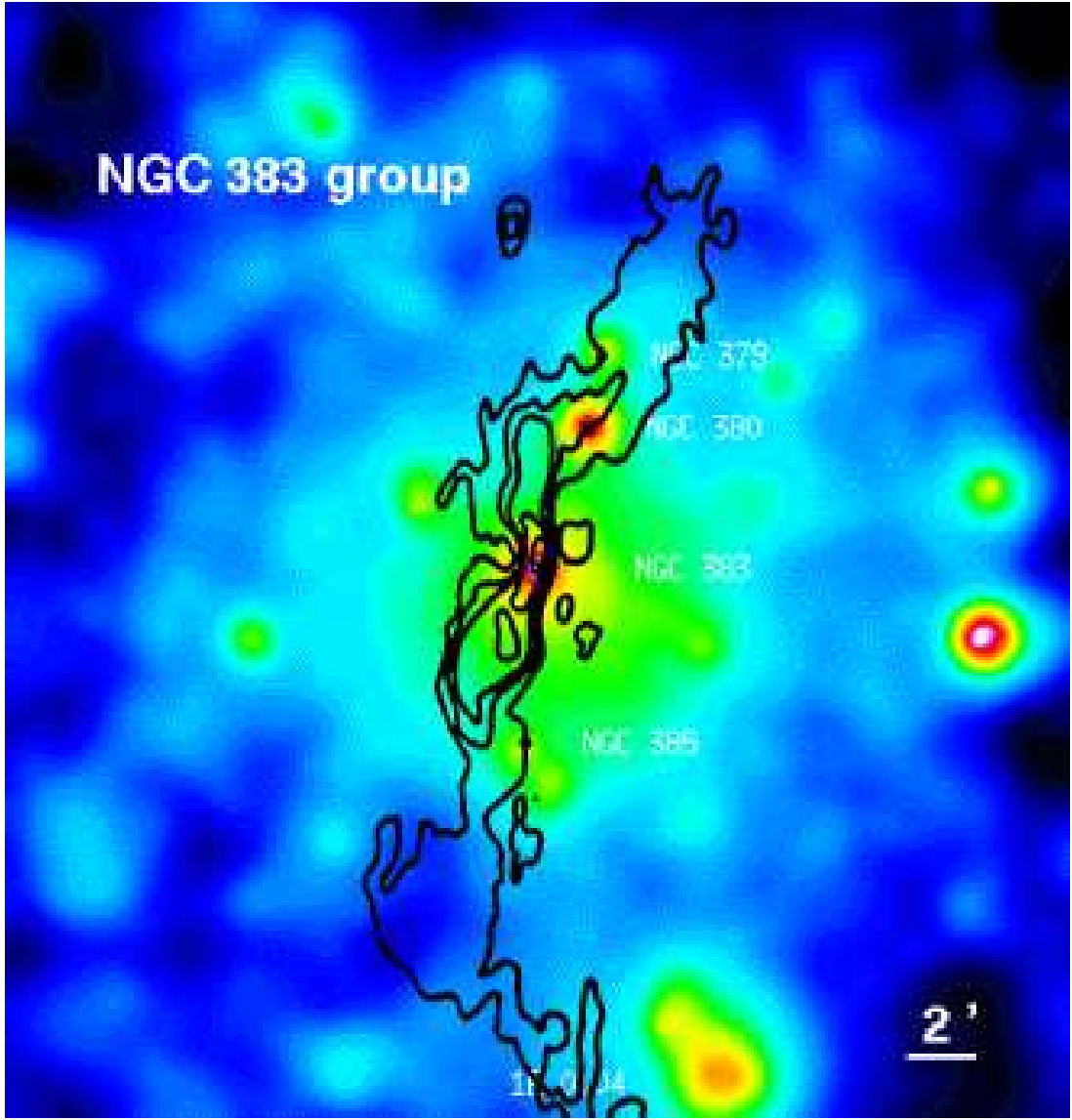}}
\vspace{-1.9cm}
\caption[dum]{ Overlay of radio jet contours {\small (Strom et al. 1983, 0.6 GHz map)}
on the X-ray image.
}
\label{dum}
\end{figure*}

  \begin{figure}[h]
\resizebox{\hsize}{!}{\includegraphics{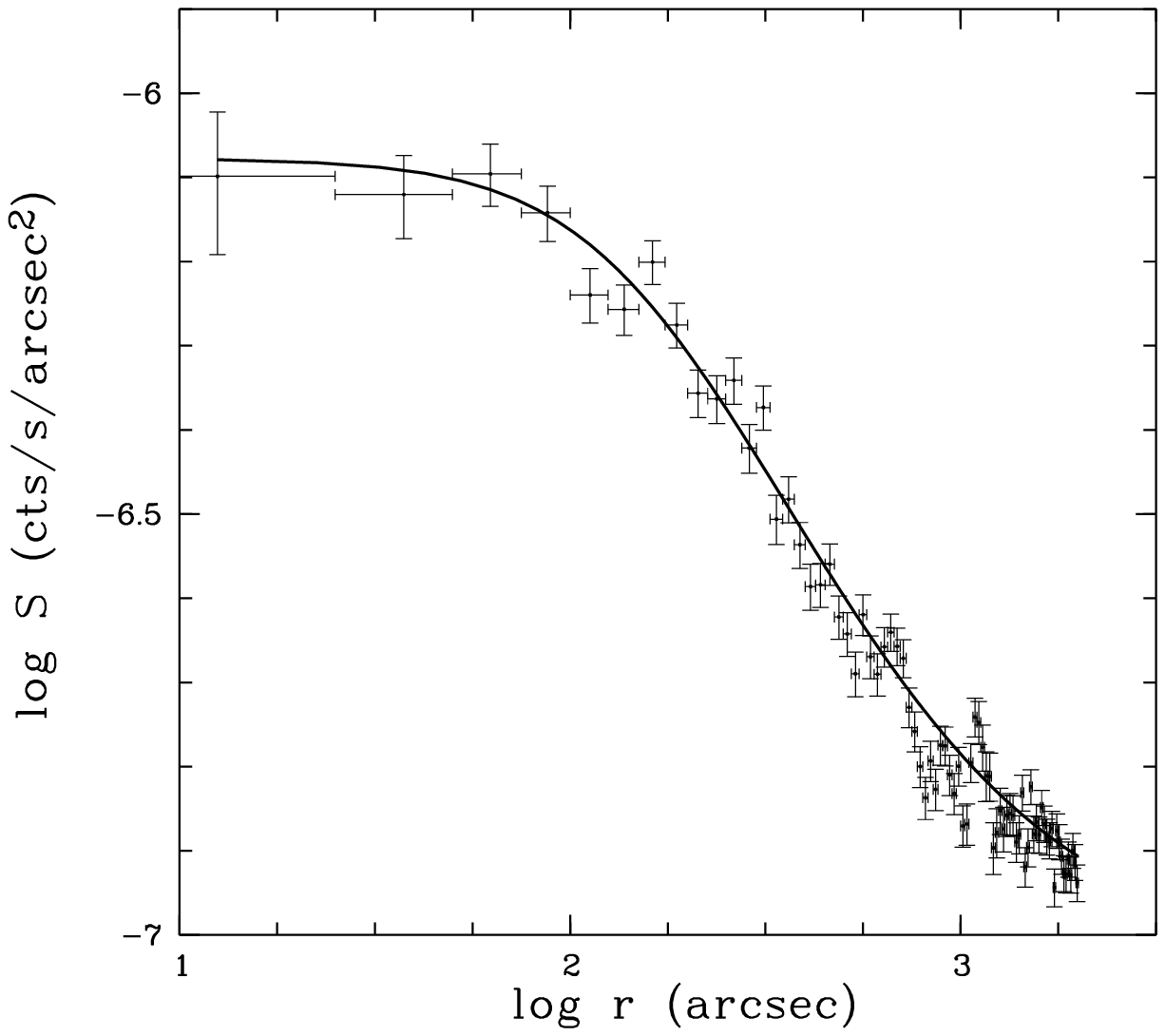}}
\vskip0.5cm
\resizebox{\hsize}{!}{\includegraphics{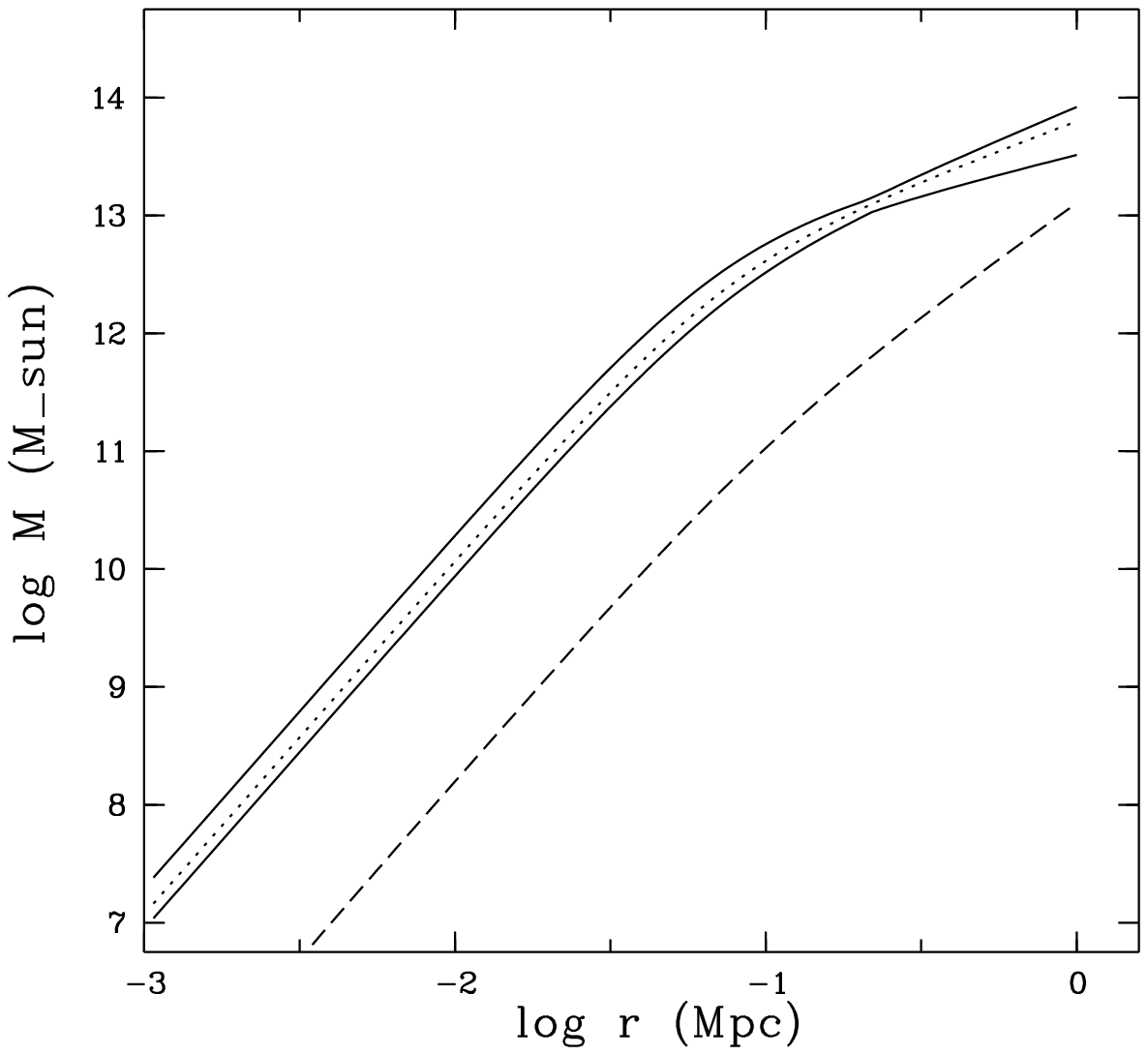}}
 \caption[profile]{{\sl Upper panel}: Observed X-ray surface brightness profile (crosses) of
 the
NGC\,383 group of galaxies {\small (excluding NGC\,383 itself)} and
best-fit $\beta$-model (solid line).
{\sl Lower panel}: Radial mass profile.
The dashed line gives the run of gas mass as obtained from the best-fit
$\beta$-model, the dotted line corresponds to the profile of the total
mass. Errors on $M_{\rm total}$ (solid lines) were derived from the temperature
range allowed by the X-ray spectral analysis, $kT$
= 1.5$\pm{0.2}$ keV,
and a temperature profile for a family of $\gamma$ models with
polytropic index $\gamma$ in the range 0.9 -- 1.3, the 
nominal temperature fixed at the core radius.}
 \label{mass}
\end{figure}

\section{X-ray properties of the central radio galaxy, 
           NGC\,383 = 3C\,31}

The optically and X-ray brightest galaxy of the group is NGC\,383.
Its spectrum is best described by a two-component model consisting of
thermal emission from a Raymond \& Smith plasma with $kT \simeq$ 0.6 keV
and a hard tail that may correspond to an AGN-like powerlaw.
The presence of the hard component can be avoided if the gas abundances
are depleted below 0.1 $\times$ solar. However, such low abundances 
are unexpected for dominant group galaxies (for a more detailed discussion
of the question of metal abundances see, e.g., Buote \& Fabian 1998).  
The amount of cold absorption is consistent
with the Galactic value in direction of NGC\,383.

The luminosity in the hard component, $L_{\rm x,h} \simeq 3.0\,10^{41}$ erg/s,
 significantly exceeds the contribution
expected from discrete stellar sources, and likely originates from the
active nucleus.  

A temporal analysis of the X-ray emission from NGC\,383 reveals 
constant source flux throughout the observation.  

The spatial position of NGC\,383 is found to be slightly off-set from
the center of the extended X-ray emission, and thus presumably from the center of
the dark matter potential.
Such off-sets of cD galaxies have also been observed in a number of other
systems. Lazzati \& Chincarini (1998) trace this back to 
a small-amplitude
oscillation of the cD galaxy around the bottom of the cluster potential.

\begin{table}[h]
 \caption{Summary of the X-ray properties of NGC\,383 (see also T97).
  $CR$ = \ros PSPC countrate. }
 \label{fitres2}
 \begin{tabular}{cccc}
 \hline
 \noalign{\smallskip}
 $CR$ & $kT$ & $L_{\rm (0.1-2.4) keV}$ & $L_{\rm B}$ \\
  \noalign{\smallskip}
  \noalign{\smallskip}
 10$^{-2}$ cts/s & keV & erg/s & erg/s \\
 \noalign{\smallskip}
 \hline
 \hline
 \noalign{\smallskip}
 2.40 & 0.6+pl & 4.7\,10$^{41}$ & 5.1\,10$^{43}$ \\
 \noalign{\smallskip}
 \hline
 \noalign{\smallskip}
 \end{tabular}
   \end{table}


\section{Radio -- X-ray relations}

\subsection {Spatial correlation ?}

NGC\,383 shows bright radio jets.
Given the spectacular examples of pressure interaction between the radio
and X-ray gas in some clusters of galaxies (e.g., Harris et al. 1994,
B\"ohringer et al. 1993, 1995), we searched for such signatures in
the NGC\,383 data. 

In the present case, we do not find conspicuous morphological
correlations between radio- and X-ray emission (Fig. \ref{dum}).
This may be partly due to the narrowness of the jet, the still limited spatial
resolution of the \ros PSPC, and the 2D view of the 3D source structure.
We note, though, that the largest extent of the southern jet transverse to
its flow direction coincides with a local minimum in the X-ray emission (Fig. \ref{dum}).  

Changes of the jet orientation angle near the locations of some optical chain galaxies,
now also detected as strong X-ray sources,
were already noted by Strom et al. (1983).

\subsection {Pressure estimates}

\begin{figure}[h] 
 \resizebox{\hsize}{!}{\includegraphics{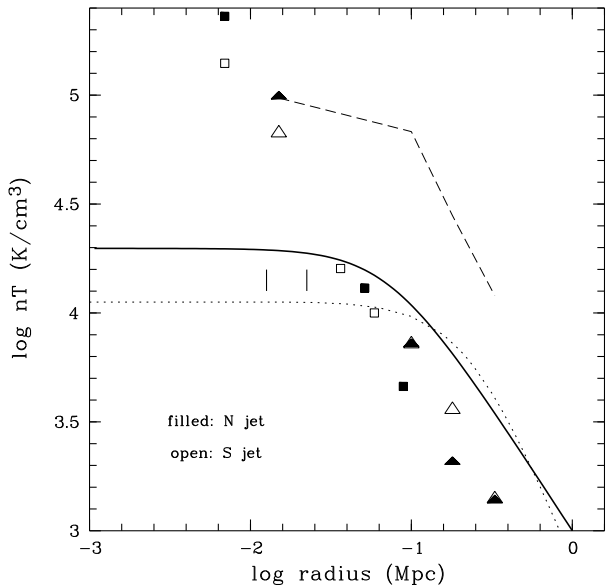}}
 \caption[pressure]{Comparison of different pressure estimates.
 The solid line gives the thermal pressure as derived from
the present \ros X-ray observation. The symbols mark the non-thermal pressure
as given in Strom et al. (1983; squares) and Morganti et al. (1988, triangles);
the filled symbols correspond to the northern jet, the open ones to the southern jet.
The vertical bars mark a scale of 25\arcsec~ (left) and the optical extent of
NGC\,383 (as given in NED; right).
Also drawn is the change in thermal pressure according to two previous
estimates (dashed line: Morganti et al. 1988; dotted: Trussoni et al. 1997).
}
 \label{pressure}
\end{figure}

The gas density and temperature derived for the X-ray gas allow a comparison
with the pressure of the radio gas, and an assessment of the confinement
of the jet material.
In Fig. \ref{pressure} we compare the pressure of the radio emitting region
as given in Strom et al. (1983) and Morganti et al. (1988){\footnote{To derive
the equipartition pressure,
they used the equations of Pacholczyk (1970) and made the standard assumptions of
equal energy density in protons and electrons, filling factor of 1, and
a powerlaw representation of the radio spectrum with cut-offs at 10 MHz and
100 GHz.}} with the thermal pressure of the X-ray gas derived from the 
density profile (Sect. 2) and a temperature of $kT$=1.5 keV.
Pressure equilibrium is reached
at about 35 kpc (projected distance from the center). Further out the thermal pressure
increasingly exceeds the nonthermal pressure.

It is interesting to note that Bridle et al. (1980)
find the expansion rate of the jets of 3C\,31 transverse to their length
to decrease with increasing distance from the radio core.
This may be related to the relative increase of the thermal pressure
of the ambient medium
with increasing radius.
While Bridle's trend refers mainly to the presently in X-rays barely resolved
core region it may be interesting to explore this relation further once 
higher-resolution X-ray data become available.

 A similar comparison of
 thermal vs. non-thermal pressure was performed in
 T97. They derived a somewhat different surface brightness profile
 and thus change in thermal pressure with the consequence that the radius
 where both pressure values are of the same order shifts further out
 which led them to suggest that NGC\,383 might have a giant halo that
 escaped detection in a (short) HRI exposure.

\section {Summarizing conclusions}
We presented results from an analysis of the
X-ray properties of the NGC\,383 galaxy group
based on \ros PSPC and HRI data.
X-ray emission can be traced out to $\sim 1 h_{50}^{-1}$ Mpc,
the estimated virial radius of the system.
We determine a total mass of $6\,10^{13} h_{50}^{-1}$
M$_{\odot}$ for the group inside this radius
with a gas mass fraction of 21\%. The intragroup
gas temperature of $1.5$ keV is both consistent
with the galaxy velocity dispersion and the
X-ray luminosity - temperature relation of groups
and clusters suggesting that the group is
fairly relaxed.

The X-ray properties of the radio galaxy NGC\,383 (3C\,31) which is located
near the center of the group were discussed, extending the work of T97.
Its spectrum is best described by a two-component model,
consisting of emission from a low-temperature Raymond-Smith plasma, and a hard tail.
The emission from NGC\,383 is not resolved by
the \ros HRI.

The possible interaction of the radio jets of 3C\,31 with the IGM was studied.
We do not find any conspicuous spatial correlation of X-ray
emission and radio jet which might be partly due to the
narrowness of the jet, the 2D view of the 3D source structure,
and the limited spatial resolution of the \ros PSPC.
With its bright extended X-ray emission, its central well-studied radio source,
and several individually detected X-ray bright member galaxies, 
the NGC\,383 group certainly is an interesting target for 
future X-ray missions like {\sl XMM} and {\sl AXAF}.

\begin{acknowledgements}
We acknowledge support from the Verbundforschung under grant No. 50\,OR\,93065.
The \ros project is supported by the German Bundes\-mini\-ste\-rium
f\"ur Bildung, Wissenschaft, Forschung und Technologie
(BMBF/DLR) and the Max-Planck-Society. \\ 
Preprints of this and related papers can be retrieved from our webpage
at http://www.xray.mpe.mpg.de/$\sim$skomossa/
\end{acknowledgements}

\end{document}